\documentclass[prl,twocolumn,nofootinbib,superscriptaddress]{revtex4}
\usepackage{graphicx,epsfig}
\usepackage{dcolumn}
\usepackage{bm}
\usepackage[american]{babel}   
\usepackage{amssymb}
\catcode`@=11
\def\citenum#1{\csname b@#1\endcsname}
\catcode`@=12
\newcommand{\kzpnn}    {\mbox{$K \! \rightarrow \! \pi \nu \overline{\nu}$ }}
\newcommand{\kpnn}    {\mbox{$K^+ \! \rightarrow \! \pi^+ \nu \overline{\nu}$ }}
\newcommand{\klpnn}   {\mbox{$K^\circ_L \! \rightarrow \! \pi^\circ \nu \overline{\nu}$ }}
\newcommand{\kpen}    {\mbox{$K^+ \! \rightarrow \! \pi^\circ e^+ \nu_e$ }}
\newcommand{\vtb}     {\mbox{$V_{tb}$ }}
\newcommand{\vts}     {\mbox{$V_{ts}$ }}
\newcommand{\vtd}     {\mbox{$V_{td}$ }}
\newcommand{\Vtd}     {\mbox{$| V_{td} |$ }}
\newcommand{\vud}     {\mbox{$V_{ud}$ }}
\newcommand{\vus}     {\mbox{$V_{us}$ }}
\newcommand{\Vus}     {\mbox{$| V_{us} |$ }}
\newcommand{\vcd}     {\mbox{$V_{cd}$ }}
\newcommand{\vcs}     {\mbox{$V_{cs}$ }}
\newcommand{\vcb}     {\mbox{$V_{cb}$ }}
\newcommand{\Vcb}     {\mbox{$| V_{cb} |$ }}
\newcommand{\vub}     {\mbox{$V_{ub}$ }}
\newcommand{\Vub}     {\mbox{$| V_{ub} |$ }}
\newcommand{\bpsiks}  {\mbox{$B^\circ_d \! \rightarrow \! J/\psi K^\circ$ }}
\newcommand{\apsiks}  {\mbox{$a_{\psi K}$ }}

\newcommand{\bdmix}   {\mbox{$\Delta M_{B_d}$ }}
\newcommand{\bsmix}   {\mbox{$\Delta M_{B_s}$ }}
\newcommand{\bsbd}    {\mbox{$\Delta M_{B_s}/\Delta M_{B_d}$ }}
\newcommand{\ek}      {\mbox{$|\varepsilon_K|$ }}
\newcommand{\Bk}      {\mbox{$\hat{B}_K$ }}
\newcommand{\sinb}    {\mbox{$\sin2\beta$ }}
\newcommand{\lamt}    {\mbox{$\lambda_t$ }}
\newcommand{\lamMS}   {\mbox{$\Lambda_{\overline{MS}}^{(4)}$ }}

\begin{document}

\title{Estimate of $B(\kzpnn)|_{SM}$ from Standard Model fits to \lamt }

\author{S.H.~Kettell}\affiliation{Brookhaven National Laboratory, Upton, New York USA}
\author{L.G.~Landsberg}\affiliation{Institute of High Energy Physics, Serpukhov, Russia}
\author{H.~Nguyen}\affiliation{Fermi National Accelerator Laboratory, Batavia, IL, USA}

\date{\today}

\begin{abstract}
We estimate $B(\kpnn)$ in the context of the Standard Model by fitting
for \lamt $\equiv V_{td}V^*_{ts}$ of the `kaon unitarity triangle'
relation. We fit data from \ek, the CP-violating parameter describing
$K$-mixing, and \apsiks, the CP-violating asymmetry in \bpsiks decays.
Our estimate is independent of the CKM matrix element \vcb and of the
ratio of B-mixing frequencies \bsbd. The measured value of $B(\kpnn)$
can be compared both to this estimate and to predictions made from \bsbd.
\end{abstract}


\maketitle


The ultra-rare FCNC kaon decays \kpnn and \klpnn are of particular
interest as these `gold-plated decays' can be predicted in the
Standard Model framework with very high theoretical accuracy.

\begin{figure}[ht]
\hskip .25in
\psfig{figure=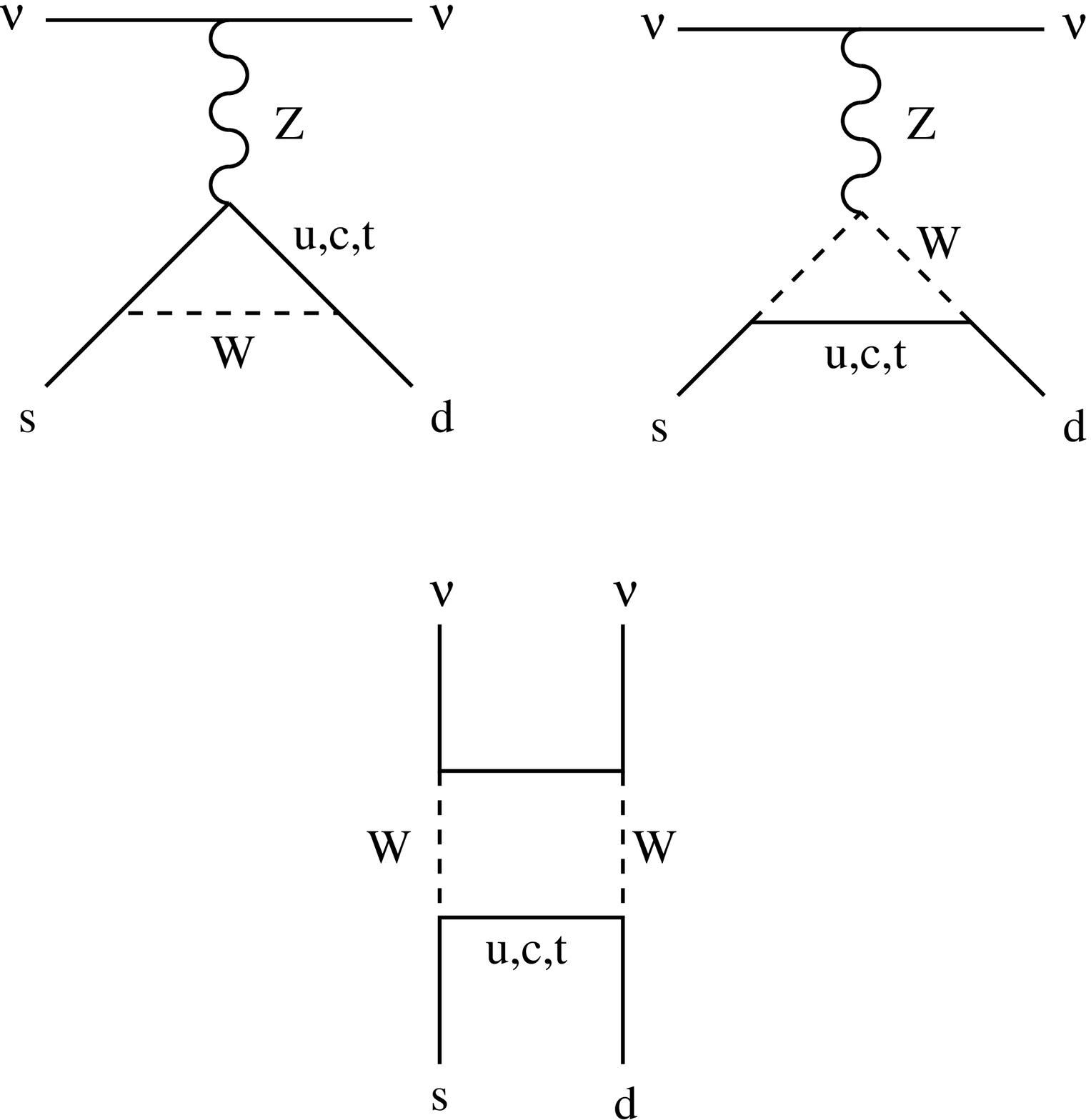,height=2.5in,width=2.5in}
\caption{The dominant contributions to \kzpnn.
\label{kpinn_beach02}}
\end{figure}

The \kzpnn decays are treated in detail in a number of
papers\cite{bb,MU,bb1,bbold,Inami-Lim,longdistance,adler,BNL949,CKM,BBL,litt,grossman,marciano,falk,blo,B_pisa,ambrosio,sa1,sa2,laplace,hocker,bk,mfv1,mfv2,B_big,BF,buras_UT,18,20,bps,22,bergmann,nir,bb2}.
We list some of the key aspects of these decays.
\begin{itemize}

\item[a)] The main contribution to these FCNC processes arises at
small distances $r\sim 1/m_t, 1/m_Z$; therefore, a very accurate
description for the strong interactions at the quark level is possible
in the framework of perturbative QCD. This analysis has been carried
out in the leading logarithmic order (LLO) with corrections to
next to leading order (NLO)\cite{bb,MU,bb1,bbold}.

\item[b)] The calculation of the matrix element $\langle\pi |
H_w|K\rangle_{\pi\nu\bar{\nu}}$ from quark-level processes involves
long-distance physics. However, these long-distance effects can be
avoided by the renormalization procedure developed by Inami and
Lim\cite{Inami-Lim}, relating the matrix element to that of the well
known decay \kpen through isotopic-spin symmetry. Other possible
long-distance contributions to $B(\kpnn)$ have been shown to be
negligible\cite{longdistance}.

\item[c)] Since the effective vertex $Zd\bar{s}$ in the diagrams of
Figure~\ref{kpinn_beach02} is short-distance, these processes are
also sensitive to the contributions from new heavy objects (e.g.,
supersymmetric particles).
\end{itemize} 

A very important step in the study of \kpnn was achieved by the E787
experiment\cite{adler} at BNL in which two clean events were found in
favorable background conditions, indicating a branching ratio of
$B(\kpnn)$ = $(15.7^{+17.5}_{-8.2}) \times 10^{-11}$.  This
observation has opened the door for future more precise study of the
\kpnn decay\cite{BNL949,CKM}.

In the Standard Model, the \kpnn decay is described by penguin and box
diagrams presented in Figure~\ref{kpinn_beach02}.  The partial widths
have the form:
\begin{eqnarray}
\Gamma(\kpnn) & = & \kappa^+ \cdot |\lambda_cF(x_c) + 
\lambda_tX(x_t)|^2  \nonumber \\
& = & \kappa^+ \cdot[ ( Re\lambda_cF(x_c) + Re\lambda_tX(x_t) )^2 \nonumber \\
& + &  (Im\lambda_cF(x_c) + Im\lambda_tX(x_t) )^2 ]  \nonumber \\
& \simeq & \kappa^+\cdot[( Re\lambda_cF(x_c)+Re\lambda_tX(x_t) )^2  \nonumber \\
& + & (Im\lambda_tX(x_t) )^2 ] \label{math/1}
\end{eqnarray}
where
\begin{eqnarray}
\kappa^+ = \left(\frac{G_F}{\sqrt{2}}\right)^2
\cdot |\langle\pi^+\nu\bar{\nu}|H_w|K^+\rangle |^2\cdot  
3\left(\frac{\alpha} {2\pi\sin^2\vartheta_w}\right)^2  \nonumber 
\end{eqnarray}
The factor of 3 in the expression for $\kappa^+$ results from the
three flavors of neutrinos $(\nu_e, \nu_\mu, \nu_r)$ participating in
the \kpnn decays.  The factors $F(x_c)$ and $X(x_t)$ are functions
corresponding to the quark loops.  These functions include the
Inami-Lim functions\cite{Inami-Lim} and the QCD corrections that have
been calculated to NLO\cite{bb,MU,bb1,bbold,BBL}.
They depend on the variables $x_i = (m_i/m_W)^2$ with the masses of
the $+\frac{2}{3}$ quarks, $m_i: i=c,t$.  The $\lambda_i \equiv
V_{id}V^*_{is}$ are vectors in the complex plane that satisfy the
unitarity relation:
\begin{equation}
\label{math/14}
\lambda_t + \lambda_c + \lambda_u = 0\quad (\lambda_i = V_{id}V^*_{is}\;;\;i=u,c,t).
\end{equation}
This equation describes the `kaon unitarity triangle', which can be
completely determined from measurement of the three kaon decays:
\kpen, \kpnn and \klpnn.  This triangle is highly elongated with a
base to height ratio of $\sim$1000.

Using the values of $m_c$ and $m_t$ in Table~\ref{tab:data}, the calculations from
Reference~\citenum{bb} yield $F(x_c)=(9.8\pm 1.8)\times10^{-4}$ and
$X(x_t) = (1.52\pm 0.05)$.  The accuracy improves with increasing quark
mass, and there are systematic dependences on \lamMS.
The $c$-quark contribution in
(\ref{math/1}) is smaller than the $t$-quark contribution, but is
non-negligible.  Although $F(x_c)/X(x_t)\sim 10^{-3}$, 
$Re\lambda_c$ is much larger than $Re\lambda_t$ and
$Im\lambda_t$. ($Re\lambda_c\sim \lambda$ while $Re\lambda_t$,
$Im\lambda_t$ and $Im\lambda_c$ are less than $\lambda^5$).

For the $CP$-violating\cite{litt,grossman} \klpnn decay
\begin{eqnarray}
\label{math/2}
\Gamma(\klpnn) & \simeq &
\frac{1}{2}|{A(K^0\to\pi^0\nu\bar{\nu}) - A(\bar{K}^0\to\pi^0\nu\bar{\nu})} |^2  \nonumber \\
& = &  \kappa^0 \cdot\frac{1}{2} | \lambda_cF(x_c) + 
 \lambda_tX(x_t) - h.c. |^2 \nonumber \\
& = &  \kappa^0 \cdot 2\left[Im\lambda_cF(x_c) + Im\lambda_tX(x_t)\right]^2 \nonumber \\
& \simeq &  \kappa^0 \cdot 2\left[Im\lambda_tX(x_t)\right]^2
\end{eqnarray}
where
\begin{eqnarray}
\kappa^0 = \left(\frac{G_F}{\sqrt{2}}\right)^2
\cdot |\langle\pi^0\nu\bar{\nu}|H_w|K^0\rangle |^2\cdot 3\left(\frac{\alpha}
{2\pi\sin^2\vartheta_w}\right)^2 \nonumber
\end{eqnarray}
The $c$-quark contribution is negligible since $Im\lambda_cF(x_c)
\ll Im\lambda_tX(x_t)$. 

The partial width for the well-known decay mode \kpen is
given by:
\begin{eqnarray}
\Gamma(\kpen) = \left(\frac{G_F}{\sqrt{2}}\right)^2
|V_{us}|^2 |\langle\pi^0e^+\nu_e|H_w|K^+\rangle |^2 \nonumber
\end{eqnarray}
As mentioned above, one can relate this to
$\langle\pi^+\nu\bar{\nu}|H_w|K^+ \rangle$ and
$\langle\pi^0\nu\bar{\nu}|H_w|K^0\rangle$ with the help of
isotopic-spin symmetry:
\begin{equation}
\left|\frac{\langle\pi^+\nu\bar{\nu}|H_w|K^+\rangle}{\langle\pi^0e^+\nu_e
|H_w|K^+\rangle}\right|^2 = \left|\frac{
\langle\pi^+|H_w|K^+\rangle}{\langle\pi^0|H_w|K^+\rangle}\right|^2 = 2r_+,
\label{math/3}
\end{equation}
\begin{equation}
\left|\frac{\langle\pi^0\nu\bar{\nu}|H_w|K^0\rangle}{\langle\pi^0e^+\nu_e
|H_w|K^+\rangle}\right|^2 = \left|\frac{
\langle\pi^0|H_w|K^0\rangle}{\langle\pi^0|H_w|K^+\rangle}\right|^2 = r_0.
\label{math/4}
\end{equation}
The factor 2 in (\ref{math/3}) accounts for the pion quark structure
$|\pi^0\rangle = \frac{1}{\sqrt{2}}|u\bar{u} - d\bar{d}\rangle$ and
$|\pi^+\rangle = |u\bar{d}\rangle$. The factors $r_+ = 0.901$ and
$r_0= 0.944$ arise from the phase space corrections and the breaking
of isotopic symmetry\cite{marciano}.

Hence from (\ref{math/1}), (\ref{math/3}) and (\ref{math/4}) the
branching ratio for the \kpnn decay is
\begin{eqnarray}
\label{math/5}
&B(\kpnn)|_{SM} = R_+\cdot \frac{X(x_t)^2}{\lambda^2} 
\nonumber \\
&  \cdot \left\{[ Re\lambda_c f \frac{F(x_c)}{X(x_t)} 
+ Re\lambda_t ]^2 + [Im\lambda_t]^2\right\}
\end{eqnarray}
where
\begin{eqnarray}
\label{math/6}
\left.{\begin{array}{rll}
R_+ & = & B(\kpen) \cdot \frac{3\alpha^2}
{ 2\pi^2\sin^4\vartheta_w}\cdot r_+  \\
 & = & 7.50\times 10^{-6} \\
f \frac{F(x_c)}{X(x_t)}  & = & (6.66 \pm 1.23)\times 10^{-4} \\ 
f & = & 1.03 \pm 0.02  \end{array}} \right\}
\end{eqnarray}
Here, $f$ is an additional correction factor to the $c$-quark term to
take into account non-perturbative effects of dimension-8
operators\cite{falk}.  The branching ratio for the \klpnn decay is
\begin{equation}
\label{math/7}
B(\klpnn)|_{SM} = R_0\cdot \frac{X(x_t)^2}{\lambda^2}[Im
\lambda_t]^2
\end{equation}
with
\begin{eqnarray}
\label{k0}
R_0  & = & R_+\cdot \displaystyle\frac{\mathstrut r_0}{r_+}\cdot  
\displaystyle\frac{\mathstrut \tau(K_L^0)}{\tau(K^+)} = 3.28
\times 10^{-5} \nonumber \\
\quad r_0/r_+ & = & 1.048 ~~~~~ \tau(K_L^0)/\tau(K^+) = 4.17 \nonumber
\end{eqnarray}

The intrinsic theoretical uncertainty of the SM prediction for
$B(\kpnn)|_{SM}$ is $\sim7\%$ and is limited by the $c$-quark
contribution, whereas for $B(\klpnn)|_{SM}$ the uncertainty is
1--2\%. However, in practice the uncertainties of the numerical evaluations of the
\kzpnn branching ratios are dominated by the current uncertainties in
the CKM matrix parameters.

The parameters $Im\lambda_t$, $Re\lambda_t$, $Re\lambda_c$ can be
estimated within the standard unitarity triangle (UT) framework using
the improved Wolfenstein parameterization\cite{blo}
$\bar{\eta},~\bar{\rho},~A,~\rm{and}~\lambda$ (with
$A\lambda^2=|V_{cb}|, \bar{\rho} \equiv \rho(1-\frac{\lambda^2}{2})$
and $\bar{\eta} \equiv \eta(1-\frac{\lambda^2}{2})$ ).  To $O(\lambda^4)$ the
CKM matrix is
\begin{eqnarray}
\label{math/10}
V_{CKM} & = &
\left( \begin{array}{ccc} 
\vud & \vus & \vub \\
\vcd & \vcs & \vcb \\
\vtd & \vts & \vtb 
\end{array} \right)  \\ \nonumber \\ 
 & = & \left(
\begin{array}{ccc}
1-\frac{\lambda^2}{2} & \lambda & A\lambda^3(\rho-i\eta) \\
-\lambda & 1-\frac{\lambda^2}{2} & A\lambda^2 \\
 A\lambda^3(1-\rho-i\eta) & -A\lambda^2 & 1
\end{array}\right) \nonumber \\ 
& + & O(\lambda^4) \nonumber 
\end{eqnarray}
and to higher order we have
\begin{equation}
\label{math/11}
\left.\begin{array}{ccl}
Re\lambda_c & = & -\lambda \left(1-\frac{\lambda^2}{2}\right)+ O(\lambda^5) \\
Re\lambda_t & = & -A^2\lambda^5\left(1-\frac{\lambda^2}{2}\right)(1-\bar{\rho})
 + O(\lambda^7) \\
Im\lambda_t & = & \eta A^2\lambda^5 + O(\lambda^9)
\end{array}\right\}
\end{equation}

The current values of these and other parameters used in this paper
can be found in Table~\ref{tab:data}.  Using (\ref{math/11}) and
Reference~\citenum{PDG} (see Table~\ref{tab:data}), equations
(\ref{math/5}) and (\ref{math/7}) can be naively solved to give the
branching ratios for \kpnn and \klpnn:
\begin{widetext}
\begin{eqnarray}
\label{math/12}
B(\kpnn)|_{SM} & = & R_+\cdot A^4\lambda^8X(x_t)^2\cdot 
\left\{\frac{1}{\sigma}[(\rho_0-\bar{\rho})^2 + (\sigma\bar{\eta})^2]\right\}
\nonumber \\
& = &  R_+\cdot |V_{cb}|^4 X(x_t)^2\cdot 
\left\{\frac{1}{\sigma}[(\rho_0-\bar{\rho})^2 + (\sigma\bar{\eta})^2]\right\} 
\nonumber \\
& = & 7.50\times 10^{-6} \cdot [2.88\times 10^{-6}\pm (19.4\%)]
[2.30 \pm (6.9\%)]\{1.44\pm  (20\%)\} \nonumber \\
& = & [7.15\pm (28.9\%)]\times 10^{-11} = [7.2\pm 2.1]\times 10^{-11}
\end{eqnarray}
\begin{eqnarray}
\label{math/13}
B(\klpnn)|_{SM} & = & R_0\cdot A^2\lambda^8X(x_t)^2\cdot
\left\{\sigma\bar{\eta}^2\right\} \nonumber \\ 
& = &  R_0\cdot |V_{cb}|^4X(x_t)^2\cdot\left\{\sigma\bar{\eta}^2\right\} 
\nonumber \\
& = & 3.28\times 10^{-5}\cdot[2.88\times 10^{-6} 
\pm (19.4\%)][2.30\pm (6.9\%)] \cdot \{ 0.129\pm    (28.6\%)\} \nonumber \\
& = & [2.8\pm (35\%)]\times 10^{-11} = [2.8\pm 1.0]\times 10^{-11}
\end{eqnarray}
\end{widetext}
with $\rho_0 = 1+\Delta = 1+ fF(x_c)/(|V_{cb}|^2X(x_t)) = 1.40\pm 
0.08$ and $\sigma = 1/(1-\frac{1}{2}\lambda^2)^2 = 1.051$.

The uncertainties of $B(\kzpnn)$ in (\ref{math/12}) and
(\ref{math/13}) are dominated by the current uncertainties in the CKM
parameters and are significantly larger
than the intrinsic theoretical uncertainties.  The uncertainty of \Vcb
is quite significant in the evaluation of $B(\kzpnn)$ due to the
$\Vcb^4$ dependence.  CLEO has recently measured\cite{21} a somewhat
higher \Vcb value of $(46.9\pm 3.0)\times 10^{-3}$, which would cause
a significant increase to B(\kzpnn) in equations (\ref{math/12}) and
(\ref{math/13}).

The numerical solutions of equations (\ref{math/12}) and
(\ref{math/13}) do not include correlations between $\bar{\rho}$,
$\bar{\eta}$, $X$ and \vcb. Rather, these calculation are used to
demonstrate the influence of different factors in the calculation of
B(\kzpnn).  An evaluation\cite{B_pisa} employing a scanning method and
conservative errors for $V_{CKM}$ obtained the following values:
$B(\kpnn)|_{SM} = (7.5 \pm 2.9) \times 10^{-11}$ and $B(\klpnn)|_{SM} = (2.6 \pm
1.2) \times 10^{-11}$.  A more recent evaluation with similar CKM
inputs, but employing a Gaussian fit obtained $B(\kpnn)|_{SM} = (7.2 \pm
2.1) \times 10^{-11}$\cite{ambrosio}. These values are not very
different from the results in equations (\ref{math/12}) and
(\ref{math/13}). In some recent analyses\cite{sa1,sa2,laplace,hocker}
with correlations included higher
precision on $B(\kzpnn)$ has been obtained.

For the values of the parameters \Vcb, $\bar{\rho}$ and $\bar{\eta}$
in equations (\ref{math/12}) and (\ref{math/13}) we adopt the more
conservative approach of Reference~\citenum{PDG}.  A more aggressive
approach\cite{bps} for the evaluation of these errors can
significantly increase the precision for $B(\kzpnn)$. Solving
equations (\ref{math/12}) and (\ref{math/13}) with these values gives
$B(\kpnn)|_{SM}=(7.4\pm1.2)\times10^{-11}$ and
$B(\klpnn)|_{SM}=(2.8\pm0.5)\times10^{-11}$.  The precision of the
outputs of the standard UT fits is dependent on the value of $\xi$,
the SU(3) breaking correction to \bsbd. The generally accepted value
of $\xi$ is $\xi = 1.15 \pm 0.06$; however, recent work would suggest
a higher value of $\xi = 1.18\pm0.04^{+0.12}_{-0.0}$\cite{lellouch}
(or even as high as $\xi = 1.32 \pm 0.10$\cite{kronfeldryan}.)

Given the strong dependence of equations (\ref{math/12}) and
(\ref{math/13}) on \Vcb, we consider an estimate of B(\kpnn) that is
essentially independent of \Vcb.  This estimate is also independent of \bsbd.  It
is based solely on \ek and \apsiks, is remarkably competitive to other
estimates, and has the advantage of simplicity.

In this work we directly evaluate \lamt to calculate $B(\kzpnn)$ from
(\ref{math/5}) and (\ref{math/7}).  This avoids the use of
$\bar{\rho}$ and $\bar{\eta}$, as has been used in previous
calculations of B(\kzpnn). This approach has been discussed in the
literature\cite{bk,22}, but as far as we know, no calculations of
$B(\kzpnn)$ exist by this method. In order to minimize uncertainty
from \Vcb, it is natural to consider \ek and \apsiks in terms of the
kaon UT\footnote{We expect that a precise determination of the apex of
the kaon UT ($\lambda_t^a$) will be available, entirely from kaon
decay data, in the near future. In the meantime, it is necessary to
use some data from the B-system, so we chose to augment \ek with the
theoretically clean measurement of the CP asymmetry \apsiks from the
B-system.}.  We recall that $\lambda_u = V_{ud}V_{us}^* \simeq
\lambda(1-\frac{1}{2}\lambda^2)$ is real, and $\lambda_c =
V_{cd}V_{cs}^*$ has a very small complex phase
$\varphi(\lambda_c)\simeq Im\lambda_t/\lambda \simeq 6\times
10^{-4}$.  The phase of \vts
is $\varphi(\vts) \simeq -\pi + Im\lambda_t*\lambda/\Vcb^2
= -\pi + 0.0172 = -\pi + 1.0^{\circ}$.  The phase of $V_{td}$ is $\varphi
(V_{td}) = -\beta$ and the angle ($\beta_K$) between $\lambda_t$ and
$\lambda_u$ is
\begin{eqnarray}
\label{math/17}
\beta_K & = & \pi  - \varphi(V_{td}V_{ts}^*) = \pi - \varphi(V_{td}) + \varphi(V_{ts}) =
\beta + 1.0^{\circ} \nonumber \\
& = & (24.6\pm 2.3)^{\circ}
\end{eqnarray}
This angle is very close to $\beta$, which in the SM is extracted
cleanly from the precise measurement of \apsiks, the CP asymmetry in
\bpsiks decays: \sinb = $0.734\pm 0.054$\cite{23}. We use an iterative
procedure, starting with $\beta_K = \beta$, from our fit to derive
$Im\lambda_t$ and recalculate $\beta_K = \beta +
Im\lambda_t*\lambda/\Vcb^2$. This procedure converges after
one iteration since the correction to $\beta$ is small. There is also
a small dependence on \Vcb; however, a 10\% change in \Vcb results in
only a 0.6\% shift in B(\kpnn), which is significantly less than the
uncertainty in our result. For all practical purposes our result is
independent of \Vcb. The preferred solution for $\beta$, based on
other SM input, such as \vub/\vcb is $\beta = (23.6\pm 2.3)^{\circ}$,
so we shall only consider this particular solution.  The extraction of
\sinb from \apsiks is also clean in models with Minimal Flavor
Violation (MFV)\cite{mfv1,mfv2,bps}. In these models there are no new
phases and all of the influences of new physics are in modifications
to the Inami-Lim functions.

\begin{figure}[ht]
\hskip .25in
\psfig{figure=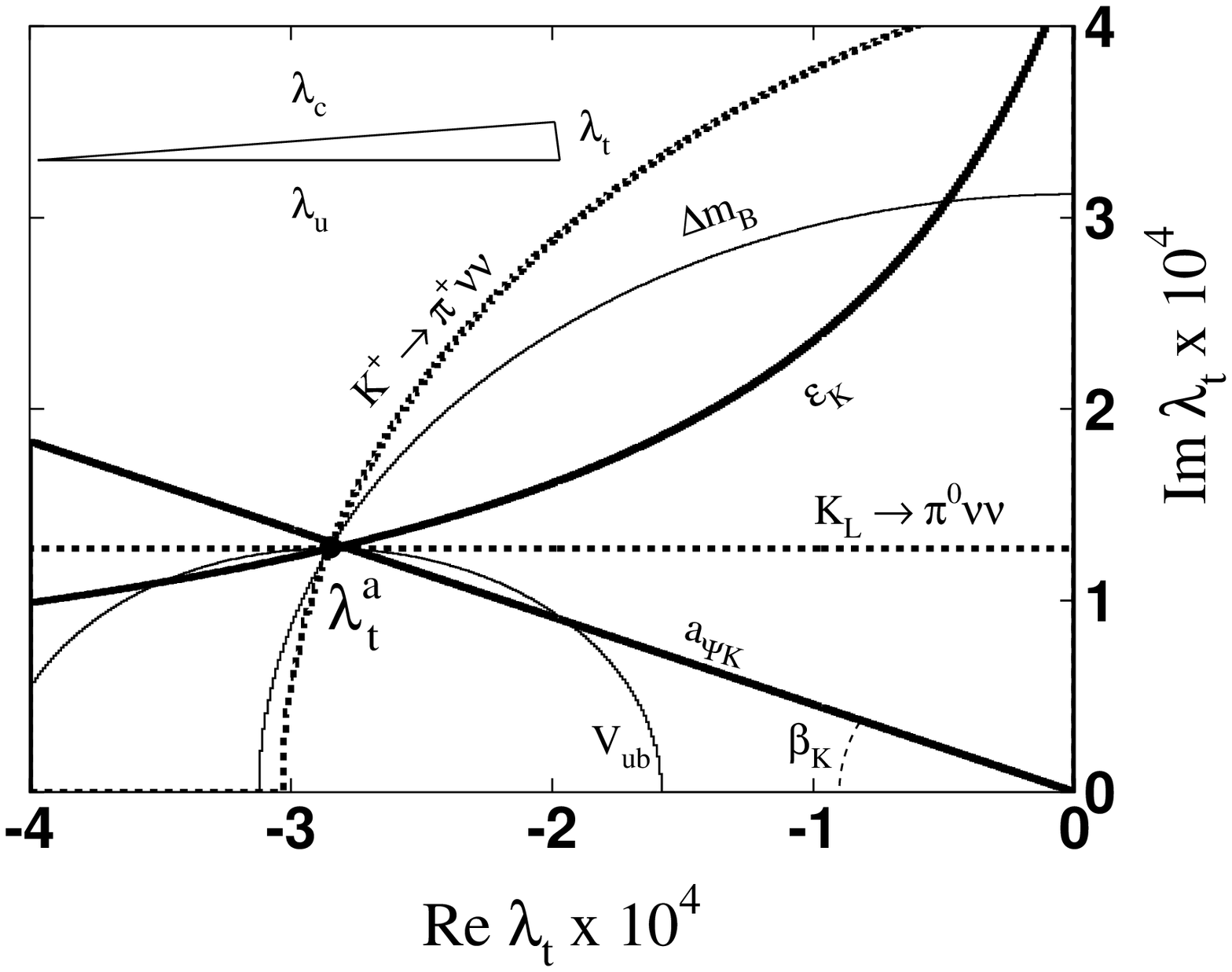,height=2.5in,width=2.5in}
\caption{The apex of the kaon unitarity triangle is $\lambda_t^a$ (no
errors are shown).  The circle labeled \vub is described by
(\ref{math/25}) with a
radius R$\sim$\vcb\vub. The thick black lines (\ek and \apsiks)
illustrate the main constraints used in this paper. The dashed lines
illustrate the constraints from \kzpnn.  The constraint from \bdmix
is shown as
the circle centered at the origin. The inset shows the triangle
(not drawn to scale).
\label{ideal_kut}}
\end{figure}

In the Standard Model, the apex of the kaon UT ($\lambda_t^a$) is
constrained by various measurements as shown in Figure~\ref{ideal_kut}
(without errors).  The constraint from \ek is expressed
as\cite{BBL,herrlich1,herrlich2,buras}
\begin{eqnarray}
\ek & = & L\cdot \Bk Im\lambda_t \cdot 
\{ Re\lambda_c[\eta_{cc} S_0(x_c) - \eta_{ct} S_0(x_c;x_t)] \nonumber \\
& & - Re\lambda_t\cdot \eta_{tt}\cdot S_0(x_t)\}
\label{math/18}
\end{eqnarray}
with parameters as shown in Table~\ref{tab:data}.  We can find the
apex of the kaon UT as the intercept of the \ek curve with
the line representing the constraint from \apsiks:
\begin{equation}
Im \lambda_t = -{\rm tan}\beta_K\cdot Re \lambda_t = 
(-0.458\pm 0.049) \cdot Re \lambda_t
\label{math/19}
\end{equation}

To calculate a probability density function (PDF) for $\lambda_t^a$,
we follow the Bayesian approach of
References~\citenum{ciuchini},~\citenum{jaffe}, and ~\citenum{bps}.
Let $f({\bf x})$ be the PDF for $\bf{x}$, where ${\bf{x}}$ is a point
in the space of ($\beta_K$, \ek, \Bk, $m_t$, $m_c$, $\lambda$,
$\alpha_s$, $\eta_{cc}$, $\eta_{ct}$, $\eta_{tt}$).  Equations
(\ref{math/18}) and (\ref{math/19}) define the mapping from ${\bf{x}}$
to $\lambda_t^a$.  Through these equations and $f({\bf{x}})$, we
derive $f(\lambda_t^a)$, the PDF for $\lambda_t^a$.  $f({\bf{x}})$
depends on the PDF's for the components of ${\bf{x}}$.  We assume that
the component PDF's are independent from one another except for the
small dependence of $\eta_{cc}$ on $m_c$ and $\alpha_s$ (discussed
below).  The component PDF's are taken from Table~\ref{tab:data}.

\begin{figure}[ht]
\hskip .25in
\psfig{figure=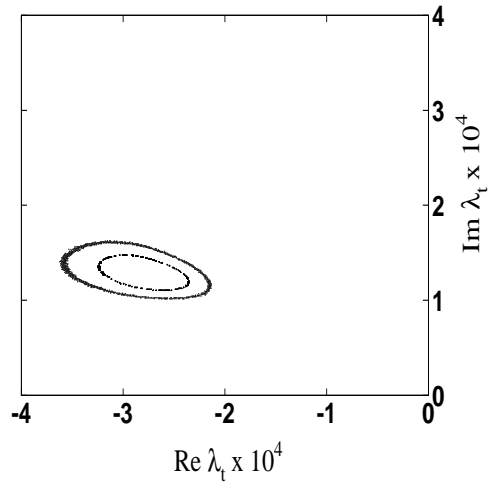,height=2.5in,width=2.5in}
\caption{1 $\sigma$ and 2 $\sigma$ C.L. intervals on $\lambda_t^a$,
obtained from the measurements of \ek and \apsiks. 
\label{limit_ek_sin2beta}}
\end{figure}

Figure~\ref{limit_ek_sin2beta} shows the PDF for $\lambda_t^a$.  We
find the following central values:
\begin{equation}
\label{math/21}
\left.\begin{array}{c}
Re \lambda_t^a = (-2.85 \pm 0.29) \times 10^{-4} \\
Im \lambda_t^a = ( 1.30 \pm 0.12) \times 10^{-4}  
\end{array}\right\}
\end{equation}

\begin{figure}[ht]
\hskip .25in
\psfig{figure=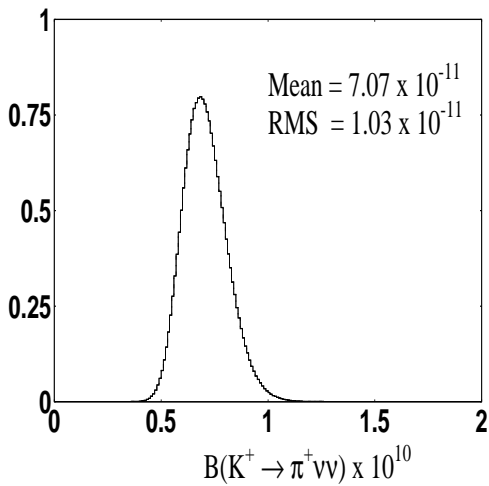,height=2.5in,width=2.5in}
\caption{The PDF for $B(\kpnn)|_{SM}$, obtained from the measurements
of \ek and \apsiks. The 95\% C.L. upper limit is $8.9 \times 10^{-11}$
and 95\% C.L. lower limit is $5.6 \times 10^{-11}$. \label{brpdf}}
\end{figure}

For $B(\kpnn)|_{SM}$ we obtain from Equations~(\ref{math/5}) and (\ref{math/21}):
\begin{eqnarray}
\label{math/22}
B(\kpnn)|_{SM} & = &\left\{\left[Re\lambda_c f F(x_c) + X(x_t)Re\lambda_t ^a\right]^2 \right. \nonumber \\
& & \left. + [X(x_t)Im \lambda_t^a]^2 \right\} \cdot \frac{R_+}{\lambda^2} \nonumber \\
& = & (7.07 \pm 1.03)\times 10^{-11}
\end{eqnarray}
The three largest contributions to the uncertainty are due to \Bk
($0.69\times10^{-11}$), $m_c$ ($0.44\times10^{-11}$) and \apsiks
($0.49\times10^{-11}$).  The probability distribution for $B(\kpnn)|_{SM}$
is presented in Figure~\ref{brpdf}.

In obtaining the results of equation (\ref{math/22}) we have accounted
for the correlations between \ek (one of the inputs for determining
$\lambda_t^a$), $F(x_c)$ and $X(x_t)$ through the variables $x_c$,
$x_t$, and \lamMS.  The functions $X(x_t)$ and $F(x_c,\lamMS)$ are
given in Reference~\citenum{bb}, from which we have parameterized
Table~1 to get:
\begin{eqnarray}
\label{math/27}
F(x_c,\lamMS)\times10^{4} & = &
9.82 + 16.58(m_c-1.3) \nonumber \\
 & & + 7.8(0.325-\lamMS)
\end{eqnarray}
where
\begin{equation}
\label{eq:MSbar}
\lamMS [GeV] = 0.341 + 16.7(-0.119 + \alpha_s(M_Z))
\end{equation}
Equation (\ref{eq:MSbar})  is accurate to 0.7\% for $\alpha_s$ in the range
0.116 to 0.122\cite{uli}. The expression for \ek (and the determination of the apex,
$\lambda_t^a$) has a dependence on $x_c$ and $x_t$ through the
Inami-Lim functions $S_0(x_c)$, $S_0(x_t)$ and $S_0(x_c,x_t)$. In
addition, the NLO correction $\eta_{cc}$ has the following
dependence\cite{uli}:
\begin{eqnarray}
\label{math/26}
\eta_{cc} & = & (1.46\pm\sigma_1)(1-1.2(\frac{m_c}{1.25} - 1)) \nonumber \\
 & & \times (1+52(\alpha_s(M_Z)-0.118))
\end{eqnarray}
with
\begin{equation}
\sigma_1 =  0.31(1-1.8(\frac{m_c}{1.25}-1))(1+80(\alpha_s(M_Z)-0.118)) 
\end{equation}
The largest correlation through $m_c$ causes both endpoints of the
vector describing B(\kpnn), $\lambda_t^a$ and $\frac{Re\lambda_c f
F(x_c)}{X(x_t)}$ to move in similar directions, so that the
uncertainty on the length of the vector is smaller than the
uncertainties in either endpoint. Inclusion of the correlations due to
$x_c$, $x_t$ and \lamMS reduces the uncertainty in $B(\kpnn)|_{SM}$ by
$\sim$20\%.

For \klpnn we obtain from (\ref{math/7}) and (\ref{math/21}):
\begin{eqnarray}
\label{math/23}
B(\klpnn)|_{SM} & = & R_0\frac{X(x_t)^2}{\lambda^2}
\left[Im\lambda_t^a\right]^2 \nonumber \\ 
& = & (2.60\pm 0.52)\times 10^{-11}
\end{eqnarray}
The four largest contributions to the uncertainty are due to \Bk 
($0.37\times10^{-11}$), \apsiks ($0.23\times10^{-11}$), $m_c$ ($0.16\times10^{-11}$) and 
$m_t$ ($0.08\times10^{-11}$).

The results of these new calculations (\ref{math/22}) and (\ref{math/23}) of
\kzpnn branching ratios from fits to \lamt are
in a good agreement with the calculations based on the standard unitarity
triangle variables (\ref{math/12}) and (\ref{math/13}) but are free of
uncertainties in \Vcb and are independent of \bsbd. 
The main source of uncertainty in (\ref{math/22}) and (\ref{math/23}) is the lattice calculation of $\Bk = 0.86\pm 0.15$.
 (We note that some lattice calculations using
domain-wall fermions\cite{cppacs,rbc,sa1} find values of \Bk
that are 10--15\% lower than the recent world average\cite{lellouch,garron}
that we use in Table~\ref{tab:data}.)  If future lattice QCD
calculations\cite{mackenzie} can significantly reduce the uncertainty in
\Bk, an improvement in $B(\kzpnn)|_{SM}$ will be
possible.

Given the difficulty of assigning PDF's to theoretical uncertainties,
we explore the influence of a more conservative scanning technique on
the uncertainty in $B(\kpnn)|_{SM}$. We determine $\lambda_t^a$ again
from only \ek and \apsiks, using gaussian errors for all quantities
except \Bk and $m_c$, which are scanned throughout their ranges:
0.72$<\Bk<$1.0 and 1.2$<m_c<$1.4.  For \Bk=0.72 and $m_c$=1.4, which
maximizes $B(\kpnn)$, the 95\% CL upper limit is
$B(\kpnn)|_{SM}<9.9\times10^{-11}$. For \Bk=1.00 and $m_c$=1.2, which
minimizes $B(\kpnn)$, the 95\% CL lower limit is
$B(\kpnn)|_{SM}>5.0\times10^{-11}$.  These limits are not much worse
than those derived from Figure~\ref{brpdf}.

\begin{figure}[ht]
\hskip .25in
\psfig{figure=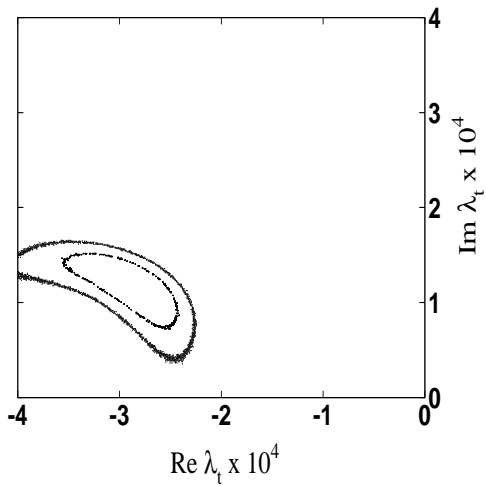,height=2.5in,width=2.5in}
\caption{1 $\sigma$ and 2 $\sigma$ C.L. intervals on $\lambda_t^a$,
obtained from the constraints of \bdmix and \Vub.
\label{Bconstraints}}
\end{figure}

\begin{figure}[ht]
\hskip .25in
\psfig{figure=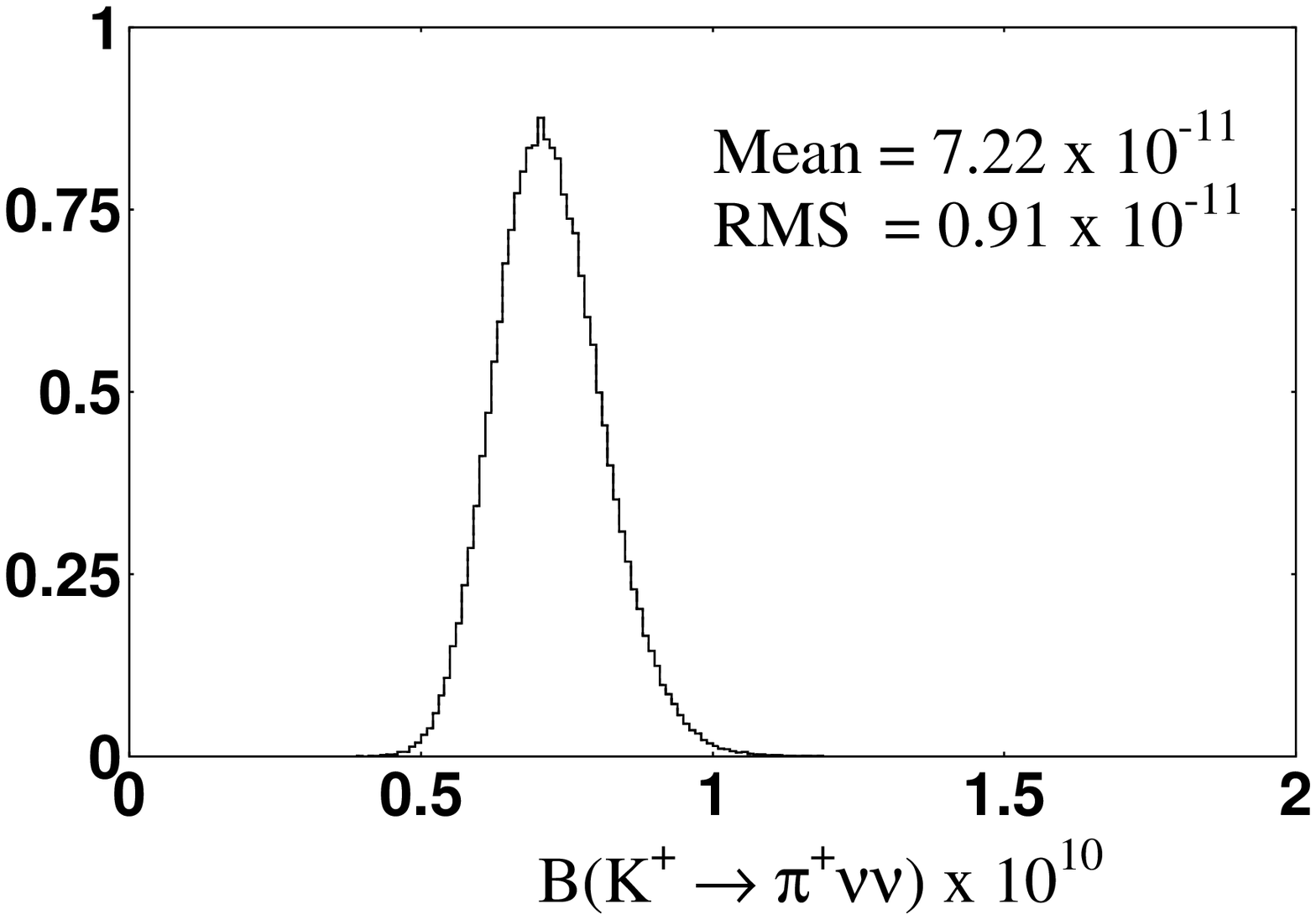,height=2.5in,width=2.5in}
\caption{The PDF for $B(\kpnn)|_{SM}$
obtained from the constraints from \ek, \apsiks, \bdmix,
and \Vub. 
\label{brcombined}}
\end{figure}

We've emphasized that our estimate uses only \apsiks and \ek.
Nevertheless, it is interesting to consider how the measurements of
\bdmix and \Vub would constrain $\lambda_t^a$. Here we will use the more
aggressive treatment of \Vcb errors (see Table~\ref{tab:data}) in order to obtain
the smallest errors on B(\kpnn). From the
following relations:
\begin{eqnarray}
\Delta m_{B_d} & = & \frac{G_F}{6\pi^2}M^2_W m_{B_d} f^2_{B_d} \hat{B}_{B_d}
\eta_{B_d} S_0(x_t) |V_{td}V^*_{tb}|^2  \nonumber \\
0 & = & V_{ud} V^*_{ub} + V_{cd}V^*_{cb} + V_{td}V^*_{tb}  \nonumber
\end{eqnarray}
and using the approximations of (\ref{math/10}): $V^*_{tb} \approx 1$, $V_{us} =
\lambda$, $V_{ud} \approx (1-\lambda^2/2)$, and $V_{cb} \approx
-V_{ts}$, we convert the equations above into:
\begin{equation}
\Delta m_{B_d} = \frac{G_F}{6\pi^2}M^2_W m_{B_d} f^2_{B_d} \hat{B}_{B_d}
\eta_{B_d} S_0(x_t) \frac{|\lambda_t|^2}{|V_{cb}|^2} \\
\label{math/24}
\end{equation}
\begin{equation}
|\lambda_t| = |V^*_{ub} V^*_{cb}(1-\lambda^2/2) - \lambda (V^*_{cb})^2 |
\label{math/25}
\end{equation}
These two equations describe two circles whose intersections contain
the apex of the kaon UT (see Fig.~\ref{ideal_kut}), and are correlated
somewhat through \vcb.  
Similar to the case of \ek, with large uncertainties from \Bk, there
are large uncertainties in the extraction of $\lambda^a_t$ from the
\bdmix and \Vub constraints, with large uncertainties from $f^2_{B_d}
\hat{B}_{B_d}$, \Vub and \Vcb.  The uncertainty on the constraint from
B-mixing may be significantly improved by the addition of \bsmix, once
the situation with $\xi$ is resolved (this will be further improved
once \bsmix is actually observed).  Using the Bayesian procedure
described earlier and the parameters in Table~\ref{tab:data}, the PDF
for $\lambda^a_t$ derived solely from the constraints of \bdmix and
\Vub is shown in Fig.~\ref{Bconstraints}.
We see that this PDF does not constrain the kaon UT apex as well as
\apsiks and \ek.  Combining all four constraints, we get the PDF
for $B(\kpnn)$ shown in Fig.~\ref{brcombined}, which is only slightly
more precise than Fig.~\ref{brpdf}.
From this combined analysis we obtain
\begin{eqnarray}
\label{math_combined}
B(\kpnn)|_{SM} & = & (7.22 \pm 0.91)\times 10^{-11} \nonumber \\
B(\klpnn)|_{SM} & = & (2.49 \pm 0.42)\times 10^{-11}.
\end{eqnarray}

The CKM matrix appears to be the dominant source of CP violation.
However, some models\cite{aguilaz} allow for a significant
contribution of new physics to $B(\kzpnn)$ while preserving the
equality between \sinb as measured from \apsiks and global CKM fits.
A crucial test of the CKM description will be to compare $\beta$
derived from $B(\kzpnn)$ to that from
\apsiks\cite{bergmann,grossman,nir,bb2}.  The most important
new information on the CKM matrix will be measurements of
$B(\kpnn)$\cite{CKM} and $B(\klpnn)$\cite{kopio} to 10\% precision.
The combination of these, in context of the SM, will determine \sinb
to 0.05\cite{B_big}, competitive with the current uncertainty on
\sinb.  The comparison of this angle obtained from B(\kzpnn) with that
from \apsiks will provide a very strong test of the SM description of
CP-violation.

Another critical test of the SM will be the direct comparison of
B(\kpnn) to either \bsbd, which in the SM both directly measure \Vtd,
or to evaluations of B(\kpnn)$|_{SM}$ such as this work.  Currently,
the E787 measurement of $B(\kpnn) = (15.7^{+17.5}_{-8.2}) \times
10^{-11}$ is consistent with the SM expectation, but the central
experimental value exceeds it by a factor of two. To date there is
only a limit on \bsmix$>14.4ps^{-1}$ (95\% C.L.)\cite{bsmixing}, but
it is likely to be observed soon.  Until \bsmix is observed, this
limit can be used to set an upper limit on $B(\kpnn)$\cite{bb}.  A
recent calculation of this limit\cite{ambrosio} gives $B(\kpnn)|_{SM}
< 13.2 \times 10^{-11}$, which is below the central experimental
value\cite{adler}.  This work used a value of $\xi = 1.15 \pm 0.06$,
whereas a higher value of $\xi$ would raise this upper limit.  Our
work is an estimation of $B(\kpnn)|_{SM}$ based solely on \ek and
\apsiks and is not dependent on \Vcb or \bsbd.  Our 95\% C.L. upper
limit is $8.9 \times 10^{-11}$ with the largest systematic error of
this approach coming from \Bk.
The uncertainty from our prediction is comparable to the expected
experimental uncertainties that might be achieved in the future
measurements of \kpnn\cite{BNL949,CKM}. An experimental measurement
significantly larger that determined from \bsbd or our 99\% C.L. limit
of $B(\kpnn)|_{SM} < 10\times10^{-11}$ will be a strong indication of new
physics.

\begin{table}[ht]
\caption{Some SM parameters used for evaluation of the standard
unitarity triangle, the kaon unitarity triangle, and
$B(\kzpnn)|_{SM}$.  The subscript G(U) denote the Gaussian(Uniform)
probability density distribution for the errors.  Errors shown without
subscripts are assumed to be Gaussian. }
\begin{tabular}[t]{l}\hline
$\left.\begin{array}{l}
\lambda = \Vus = 0.222\pm 0.002 \\
\end{array}\right.\begin{array}{l}
{\rm ~ } \end{array}$ \\
$\left.\begin{array}{l}
\bar{\rho} = 0.22\pm 0.10 \\
\bar{\eta} = 0.35\pm 0.05 \\
|V_{cb}| = (41.2\pm 2.0)\cdot 10^{-3} \\
\Vub = (3.6\pm0.7)\times10^{-3}\\
\end{array}\right\}\begin{array}{l}
${\rm PDG---2002~\cite{PDG} }$ \end{array}$ \\
$\left.\begin{array}{l}
\bar{\rho} = 0.173\pm 0.046 \\
\bar{\eta} = 0.357\pm 0.027 \\
|V_{cb}| = (40.6\pm 0.8)\cdot 10^{-3} \hspace{1cm} \\
\end{array}\right\}\begin{array}{l}
${\rm \cite{bps} }$ \end{array}$ \\
$\left.\begin{array}{l}
\beta_K = \beta + 1^\circ = (24.6 \pm 2.3)^\circ \\
\ek = (2.282 \pm 0.017) \cdot 10^{-3}$~\cite{PDG}$  \\
\Bk = 0.86 \pm 0.06_{\rm G} \pm 0.14_{\rm U}$~\cite{lellouch,ckmutws}$ \\
m_c = \bar{m}_c = 1.3\pm 0.1 {\rm GeV/c^2} \\
\end{array}\right.\begin{array}{l}
{\rm  } \end{array}$ \\
$\left.\begin{array}{l}
m_t = \bar{m}_t = 166\pm 5 {\rm GeV/c^2} \\
X(x_t) = 1.52\pm 0.05 \\
F(x_c) = \frac{2}{3}X^e_{NL}(x_c)+\frac{1}{3}X^\tau_{NL}(x_c) \\
 = (9.82 \pm 1.78) \cdot 10^{-4} \\
\lamMS = 0.325 \pm 0.08  {\rm GeV} \\
\end{array}\right\}\begin{array}{l}
${\rm ~\cite{bb} }$ \end{array}$ \\
$\left.\begin{array}{l}
f = 1.03 \pm 0.02$~\cite{falk}$ \\
f \cdot F(x_c)/X(x_t) = (6.66 \pm 1.23) \cdot 10^{-4}
\end{array}\right.\begin{array}{l}
{\rm ~ } \end{array}$ \\
$\left.\begin{array}{l}
S_0(x_c) = (2.42\pm 0.39)\cdot 10^{-4} \\
S_0(x_c,x_t) = (2.15\pm 0.31)\cdot 10^{-3} \\
S_0(x_t) = 2.38\pm 0.11 \\
\eta_{cc} = 1.45\pm 0.38$~\cite{herrlich1}$ \\
\eta_{ct} = 0.47\pm 0.04$~\cite{herrlich2}$ \\
\eta_{tt} = 0.57\pm 0.01$~\cite{buras}$ \\
L = 3.837\times10^{4}$~\cite{B_big}$
\end{array}\right.\left\}\begin{array}{l}
{\rm ~}\\
{\rm Inami-Lim}\\
{\rm functions \ and} \\
{\rm QCD \ corrections} \\
{\rm for}\,K^0\rightleftarrows \bar{K}^0\,{\rm and} \\
\ek\,{\rm evaluation}\\
{\rm ~}
\end{array}
\right.$
\\
$\left.\begin{array}{l}
\Vcb({\rm incl.})=(40.4\pm0.7_{\rm G}\pm0.8_{\rm U})\cdot 10^{-3}$~\cite{artuso_barberio}$ \\
\Vcb({\rm excl.})=(42.1\pm1.1_{\rm G}\pm1.9_{\rm U})\cdot 10^{-3}$~\cite{artuso_barberio}$ \\
\Vub({\rm incl.})=(40.9\pm4.6_{\rm G}\pm3.6_{\rm U})\cdot 10^{-4}$~\cite{bps}$ \\
\Vub({\rm excl.})=(32.5\pm2.9_{\rm G}\pm5.5_{\rm U})\cdot 10^{-4}$~\cite{bps}$ \\
\Delta m_{B_d} = 0.489 \pm 0.008 ~{\rm ps^{-1}}$~\cite{PDG}$  \\
f_{B_d} \sqrt{\hat{B}_{B_d}} = 230 \pm 30_{\rm G} \pm 15_{\rm U}~{\rm MeV}
\end{array}\right.$
$\left\}\begin{array}{l}
\bdmix \\
{\rm and} ~ \Vub \\
{\rm parameters}\\
{\rm used ~ in}\\
{\rm evaluating~the}\\
{\rm constraint ~ on } \\
{\rm \lambda_t^a ~ in ~ Fig.~\ref{Bconstraints} } 
\end{array}
\right.$\\
$\left.\begin{array}{l}
\xi = \frac{f_s}{f_d}\sqrt{\frac{\hat{B}_s}{\hat{B}_d}} =1.15 \pm 0.06\\
\end{array}\right.
\left\}\begin{array}{l}
{\rm old ~ value}\\
\end{array}
\right.$\\
$\left.\begin{array}{l}
\xi = 1.32\pm0.10$~\cite{kronfeldryan}$\\
\xi = 1.18\pm0.04^{+0.12}_{-0.0}$~\cite{lellouch}$\\
\xi = 1.22\pm0.07$~\cite{becizevic}$\\
\end{array}\right.
\left\}\begin{array}{l}
{\rm  ~ }\\
{\rm new ~ data ~ with}\\
{\rm chiral ~ log ~ extrapolation}\\
\end{array}
\right.$\\

\label{tab:data}
\end{tabular}
\end{table}

\section{Acknowledgments}
We would like to thank P.~Cooper, G.~Isidori, D.E.~Jaffe, P.~Mackenzie, 
U.~Nierste, L.~Okun and A.~Soni
for useful discussions and comments and W.~Marciano for the original
stimulation to perform this type of calculation. This work was in part
supported by the U.S. Department of Energy through contract
\#DE-AC02-98CH10886, in part by the Fermilab Particle Physics
Division and in part by IHEP. L.G.L is grateful to the Fermilab Administration and
Particle Physics Division for their hospitality and support for his stay
at Fermilab during the preparation of this work.

\section{Note}

During the final preparation of this work for publication we found
that Reference~\citenum{stocchi} considered fitting for the apex of
the UT from the CP-violating data only (\ek and \apsiks), as we do.
However, Reference~\citenum{stocchi} used ($\bar{\rho},\bar{\eta}$),
which is dependent on \Vcb and is not as suitable for analysis of
\kzpnn.

\end{document}